\documentclass[aps,twocolumn,groupedaddress,showpacs,floats,amssymb]{revtex4}
\usepackage[dvips]{graphicx}
\usepackage{bm}
\begin{document}

\title{Two-Dimensional Weakly Interacting Bose Gas in the
Fluctuation Region }

\author{Nikolay Prokof'ev}
\author{Boris Svistunov}

\affiliation{Department of Physics, University of
             Massachusetts, Amherst, MA 01003, USA}
\affiliation{Russian Research Center ``Kurchatov Institute",
             123182 Moscow, Russia}

\date{\today}

\begin{abstract}
We study the crossover between the mean-field and critical
behavior of the two-dimensional Bose gas throughout the
fluctuation region of the Berezinskii--Kosterlitz--Thouless phase
transition point. We argue that this crossover is described by
universal (for all weakly interacting $|\psi|^4$ models )
relations between thermodynamic parameters of the system,
including superfluid and quasi-condensate densities. We establish
these relations with high-precision Monte Carlo simulations of the
classical $|\psi|^4$ model on a lattice, and check their
asymptotic forms against analytic expressions derived on the basis
of the mean-field theory.
\end{abstract}


\maketitle

\section{Introduction}
\label{sec:introduction}
The mean-field (MF) approach to the weakly interacting Bose gas
(BG) is a well established theoretical tool
\cite{bogoliubov,LP,Popov}. However, it is not adequate in the
fluctuation region of the  superfluid phase transition. The
situation is most dramatic in the two-dimensional (2D) case, where
the size of the fluctuation region, $\Delta T$, is almost insensitive
to the smallness of interaction \cite{FH} ($\hbar =1$):
\begin{equation}
\Delta T/T_c \sim 1/\ln(1/mU) \; . \label{fl_region}
\end{equation}
Here $T_c$ is the critical temperature, $m$ is the particle mass,
and $U$ is the effective long-wavelength interaction constant. The
regime of weak interaction corresponds to a small dimensionless
parameter
\begin{equation}
mU \ll 1 \; , \label{mU}
\end{equation}
which close to the transition point is equivalent to the condition
$ nU \ll T $ ($n$ is the particle density).

For the purposes of the present paper the microscopic origin of
the effective interaction $U$ is not important. However, to make
connection between our results and realistic experimental systems
we briefly review how $U$ relates to the interatomic interaction
potential $V(r)$. The value of $U$ corresponds to the pair vertex
(the sum of ladder diagrams) with typical external momenta $\sim
n^{1/2}$. It can be written in a generic form as
\begin{equation}
U={V_0 \over 1+(mV_0/4\pi ) \ln (1/nd^2) } \; , \label{U-V}
\end{equation}
where microscopic parameters $V_0$ and $d$ depend on the case. The
simplest one is that of a weak short-ranged potential satisfying
the condition $V(r \sim r_0) \ll 1/mr_0^2$, where $r_0$ is the
potential radius. In this case, $V_0=\int V(r) \, d^2 r$, and with
logarithmic accuracy assumed in Eq.~(\ref{U-V}) one may put
$d=r_0$ provided $nr_0^2 \ll 1$ (when $nr_0^2 \gtrsim 1$ the
logarithmic term in the denominator can be neglected). In the {\it
quasi}-2D system \cite{KSS,Petrov}, when the localization length
of 3D atoms in the direction perpendicular to the 2D plane (axis
$\hat{\bf z}$) is much larger than $r_0$, one has first to average
the pair interaction over a wavefunction in the $\hat{\bf
z}$-direction, $\phi_0(z)$. Now, $V_0=(4\pi a/m) \int
|\phi_0(z)|^4 dz$, where $a$ is the 3D scattering length.
Introducing the localization length by $l^{-1}_z=\int
|\phi_0(z)|^4 dz$, we have $V_0=4\pi a/ml_z$. With the same
logarithmic accuracy, in Eq.~(\ref{U-V}) $d \approx l_z$ for
$nl_z^2 \ll 1$ (cf. Refs.~\onlinecite{KSS,Petrov}). Finally, the
case of strong short-ranged potential [which in the context of
weakly interacting gas implies $\ln (1/nr_0^2) \gg 1$] formally
corresponds to the limit $V_0 \to \infty$ in Eq.~(\ref{U-V}). In
this case the effective interaction depends only on the parameter
$\ln (1/nd^2)$ with $d \approx r_0$ (see
Ref.~\onlinecite{Schick}).

Because of weak log-dependence of the fluctuation region on
interaction one may wonder whether the MF theory makes sense at
all in 2D (apart from the academic limit of exponentially small
$mU$), and, if it does, then when. As a characteristic example of
how problematic it is to reach the proper asymptotic limit,
consider a dilute gas with very small $nd^2$ when $mU \sim 4\pi
/\ln (1/nd^2) \ll 1$. With the same logarithmic accuracy it
follows then that the critical point can be found without even
resorting to the Berezinskii--Kosterlitz--Thouless physics
\cite{Popov,FH}, $T_c \approx  (2\pi n / m)  \ln^{-1} (1/mU)$.
Meanwhile, a more accurate result for the critical point is (for
future reference and convenience we write the answer for critical
density as a function of temperature) \cite{PRS} :
\begin{equation}
n_c = {mT \over 2 \pi} \ln \left( { \xi \over mU } \right) \; ,
~~~~\xi=380\pm 3 \; . \label{xi_n}
\end{equation}
Obviously, an enormous value of $\xi$ makes it virtually impossible
to reach the limit of small $U$ when $\xi$ can be ignored.

There is, however, a very important point about the fluctuation
region of a weakly interacting BG: In the limit of small $U$ all
$|\psi|^4$ models---quantum or classical, continuous or
discrete---allow a universal description \cite{Popov,Baym}. This
observation follows from a simple fact that interactions are
important only for long-wavelength components of the order
parameter field, $\psi ({\bf r} )$, with momenta $k \lesssim k_c =
m \sqrt{UT} \ll k_T = \sqrt{mT}$, and in this limit the effective
Hamiltonian is given by the $|\psi |^4$ model
\begin{equation}
H[\psi]= \int \left\{  {1 \over 2m}|\nabla \psi|^2 + {U \over 2}
|\psi|^4 - \mu |\psi|^2 \right\} \, d {\bf r} \; ,
\label{psi4}
\end{equation}
where $\mu $ is the effective chemical potential. The microscopic
physics of the model is important only at much higher momenta, $k
\gg k_c$,  where the system behavior is ideal (in linear in $U$
approximation) and thus may be easily accounted for analytically.

This observation was successfully used in
Refs.~\onlinecite{KPS,AM,PRS} (both in 3D and 2D) in the study of
the critical point dependence on interaction. Same considerations
apply, though, not only to the critical point itself, but to the
whole fluctuation region around it, and one thus expects that,
e.g., the superfluid density dependence on density, $n_s (n-n_c)$,
or chemical potential, $n_s(\mu -\mu_c)$, is also universal close
to the transition point and into the region where the MF theory
takes over.

The study of this universal behavior is the subject of this paper.
We found that even for very weak interaction, say,  $ mU \sim
0.01$, the conventional MF theory result $n_s/n = 1 -T/T_c$ may
not be used since the fluctuation region is still of order $T_c$
itself. However, if $T_c$ is related to the density by
Eq.~(\ref{xi_n}), the modified version of the MF theory developed
in this paper works remarkably well. In particular, the Equation
of state and the quasicondensate density may be predicted very
accurately up to $T_c$ (this is not true for the superfluid
density, though).

In Sec.~\ref{sec:relations} we establish the universal form of the
Equation of state and the dependence of the superfluid density and
quasicondensate density on chemical potential along with their
asymptotic behavior away from the critical point. In
Sec.~\ref{sec:numerical_procedure} we describe the numeric model
and the simulation procedure. Our results are presented and
compared to the MF and Kosterlitz--Thouless theories in
Sec.~\ref{sec:comparison}. We conclude in
Sec.~\ref{sec:conclusion} with discussing the obtained universal
results in the context of quantum Bose gases.
\section{Universal relations for weakly
interacting $|\psi|^4$ models} \label{sec:relations}

The critical point of the BG is defined by Eq.~(\ref{xi_n})
and the corresponding relation for the chemical potential
\cite{PRS}
\begin{equation}
\mu_c = {mTU \over \pi} \ln \left( { \xi_\mu \over mU } \right) \;,
~~~~\xi_\mu=13.2\pm 0.4
\label{xi_mu}
\end{equation}
which can be rewritten in the form
\begin{equation}
\mu_c = 2n_cU + {mTU \over \pi} \ln \left( { \xi_\mu \over \xi } \right) \;.
\label{xi_mu2}
\end{equation}
Both $n_c$ and $\mu_c$ are model specific, and their values depend
on the ultra-violet cutoff, $k_*$, through the logarithm $\ln
(k_{*}/k_c)$; for the quantum gas $k_* \sim k_T$. However, if the
dominant MF type contribution to the chemical potential, $2nU$, is
subtracted, the difference is an ultra-violet-cutoff-independent
quantity. The same is true for the difference $\mu -\mu_c$ or
$n-n_c$. It seems natural then to introduce a dimensionless
variable
\begin{equation}
 X=(\mu -\mu_c)/mTU \; ,
 \label{X}
\end{equation}
as a universal control parameter with the typical variation
across the fluctuation region of order unity. The Equation of state
may then be written in the universal form as
\begin{equation}
 {2nU-\mu \over mTU} = \theta (X) \; ,
 \label{theta}
\end{equation}
where  $\theta$ is a dimensionless function. By subtracting
critical values from $n$ and $\mu$ we can restate it as
\begin{equation}
n-n_c = mT \lambda(X) \; , ~~~~~  \lambda(X) = [\theta (X) -
\theta_0 + X]/2 \; ,
 \label{dn}
\end{equation}
with
\begin{equation}
\theta_0 \equiv \theta(0) = {1 \over \pi }~\ln (\xi /\xi_{\mu}) \;
.
 \label{rel1}
\end{equation}
From previous results \cite{PRS} we have $\theta_0 = 1.07 \pm
0.01$, with the error bar being largely determined by the
uncertainty in $\xi_\mu$. The $\lambda(X)$ function describes the
so-called adsorption isotherm which is relevant to the situation
where the 2D system is formed by atoms adsorbed on a surface. In
the case of a trapped gas, the function $\lambda(X)$ describes the
density profile of the gas in the hydrostatic regime (see the
discussion in Sec.\ \ref{sec:conclusion}).

The behavior of the superfluid density is described by a
dimensionless function $f$:
\begin{equation}
n_s =  (2mT /\pi )  \, f(X) \; .
\label{f}
\end{equation}
According to the Kosterlitz-Thouless theory \cite{KT}
(see also below),
\begin{equation}
f(X\to +0 ) \longrightarrow 1 + \sqrt{2 \kappa ' X} \; ,
\label{K}
\end{equation}
where $\kappa '$ is some constant to be defined numerically.

Finally, in the superfluid region an important quantity is the
quasicondensate density $n_0$, which we define by the relation
\begin{equation}
n_0=\sqrt{Q} \; , ~~~Q =  2\langle \, |\psi|^2\rangle^2  - \langle
\, |\psi|^4 \rangle \;.
\label{Q}
\end{equation}
The idea behind this definition is as follows. The notion of
the quasicondensate \cite{KSS} implies that the field
$\psi$ has the following structure
\begin{eqnarray}
\psi ({\bf r})  & = & \psi_0 ({\bf r}) + \psi_1 ({\bf r}) \; ,
\label{psi} \\ \psi_0 ({\bf r})& \approx &  \sqrt{n_0} \, e^{i
\Phi ({\bf r})} \; , \label{psi_0}
\end{eqnarray}
where $n_0$ is called the quasicondensate density, and $\psi_1$ is
the Gaussian field independent of $\psi_0$. Under these
conditions, $Q\equiv n_0^2$. [The Gaussian field $\psi_1$ obeys
the Wick's theorem and thus does not contribute to $Q$.]  Away
from the superfluid region the notion of quasicondensate gradually
becomes ill-defined, but the quantity $Q$ is still of interest as
a measure of local non-Gaussian correlations. We will jargonically
use the term 'quasicondensate density' even well inside the
fluctuation region, understanding by $n_0$ the quantity
$\sqrt{Q}$.

Since the MF theory result predicts $n_0\approx n_s$, it is
appropriate to characterize the dependence of $Q$  on $\mu$ in
close analogy to Eq.~(\ref{f}):
\begin{equation}
\sqrt{Q} = ( 2mT / \pi) \, g(X) \; . \label{g}
\end{equation}
The three functions --- $~\theta(X)$, $f(X)$, and
$g(X)~$ --- completely characterize system properties in the
vicinity of the critical point

We now turn to the MF and Kosterlitz--Thouless theories to
establish asymptotic behavior of functions $\theta(X)$, $f(X)$,
and $g(X)$.

\noindent \underline{Asymptotic behavior at $X \to \infty$}.
The notion of quasicondensate is well-defined in this region and
its density $n_0$ obeys a typical MF relation \cite{KSS}
\begin{equation}
 (n_0+2n')U = \mu \; ,
\label{rel0}
\end{equation}
where
\begin{equation}
n'=n-n_0
\label{n_0}
\end{equation}
is the non-quasicondensate part of the particle density. [One may worry
how far the analogy between the genuine condensate and
quasicondensate goes; the answer is that apart from the long-range
order problem they are indistinguishable at the MF level, and our
simulations confirm this assertion.] Comparing
(\ref{rel0})-(\ref{n_0}) with (\ref{theta}), we see that
\begin{equation}
n_0(X \to \infty ) \longrightarrow
 \, mT \, \theta(X)\; .
\label{rel3}
\end{equation}
or
\begin{equation}
g(X \to \infty ) \longrightarrow (\pi/2) \, \theta (X) \;.
\label{lim2}
\end{equation}

An explicit expression for the non-quasicondensate part is given by
\cite{rem1}:
\begin{equation}
n'=\int \frac{{\rm d}^2 k}{(2\pi )^2}
 \left[ \frac{\epsilon (k) + n_0 U -E(k)}{2E(k)}
+ \frac{\epsilon (k) \, \nu_E}{E(k)} \right] \; ,
\label{rel7}
\end{equation}
where $\epsilon(k)=k^2/2m$ is the free particle dispersion law,
$E(k)= \sqrt{\epsilon (k)[\epsilon (k) + 2n_0 U]}$ is the
Bogoliubov quasiparticle spectrum, and $\nu_E =[ \exp[E(k)/T]-1 ]
^{-1}$ is the Bose distribution function. In the region of
interest, $\theta \sim 1$, the first term in the integral is
smaller than the second one by the gas parameter $mU \ll 1$, and
should be omitted \cite{rem}. With the same accuracy, the second
term yields
\begin{equation}
 n'\approx -(mT/2\pi) \ln (2n_0U/T) \; .
\label{rel8}
\end{equation}
With the help of Eq.~(\ref{rel3}) the total density $n=n'+n_0$
may be now written as
\begin{eqnarray}
 n &\approx &{mT \over 2\pi} \ln (1/mU)+ {mT \over 2}
 \left[2\theta -{1\over \pi} \ln (2\theta )  \right] \nonumber \\
   &\equiv & n_c +
 {mT \over 2} \left[2\theta   - {1\over \pi}
 \ln (2 \xi \theta )  \right]\; .
 \label{rel8b}
\end{eqnarray}
Substituting this relation into (\ref{dn})
we find the asymptotic behavior of $\theta (X)$:
\begin{equation}
\theta - \pi^{-1} \ln \theta \, \to \,  X + \pi^{-1} \ln
(2\xi_{\mu})~~~{\rm at}~~X \to \infty \; . \label{lim1}
\end{equation}

To find $f(X\to \infty )$, we consider the
standard expression for the normal component density  \cite{LP}
\begin{equation}
 n_n = -{1 \over 2m} \int \frac{{\rm d}^2 k}{(2\pi )^2} \,
 \left[ {{\rm d  \nu_E} \over {\rm d E}} \right] \, k^2 \; ,
\label{normal}
\end{equation}
and compare it to the expression (\ref{rel7}) for the non-quasicondensate part
of the particle density. After integration by parts and
a straightforward algebra we find that up to higher-order in $mU$
terms, $n_n-n'=mT/2\pi$, which means that
\begin{equation}
 n_s = n_0 - mT/2\pi\; ,
\label{ns_n0}
\end{equation}
and, accordingly,
\begin{equation}
 f(X \to \infty ) \longrightarrow \,  g(X) - 1/4
 \longrightarrow (\pi/2) \, \theta (X) -1/4 \; .
\label{f_g}
\end{equation}

It is important to note that while for obtaining asymptotic
relations (\ref{lim1}) and (\ref{ns_n0})-(\ref{f_g}) we employed
the theory of the weakly interacting {\it quantum Bose gas}, the
final results are valid for {\it any} weakly interacting 2D
system of the $|\psi|^4$ universality class, since these
pertain to the universal long-wave behavior of the system.

\noindent \underline{Asymptotic behavior at $X \to - \infty$}.
In the region $X<0$ the $f$-function is identically zero; the
quasicondensate density is of no special interest in the normal
phase [$g(X)\to 0$ in this limit]. Hence, the only quantity we
have to look at is the Equation of state,  $\theta (X)$. Once
again, we resort to the MF equation for the effective chemical
potential $\mu ' = \mu - 2nU=-\theta mUT   $, and calculate the
total density from the integral:
\begin{eqnarray}
n & \approx & \int \frac{{\rm d}^2 k}{(2\pi )^2} \, \left[ \exp (\epsilon/T
+ \theta mU) - 1 \right]^{-1}
\label{rel9} \\
  & \approx &  - (mT/2\pi)   \ln (\theta m U) \equiv
 n_c -{mT \over 2\pi } \ln (\theta \xi)  \; .
 \label{rel10}
\end{eqnarray}
This expression may be immediately related back to the Equation of
state (\ref{theta}) and leads to the relation
\begin{equation}
\theta + \pi^{-1} \ln \theta \, \to \,  |X| -\pi^{-1} \ln
\xi_{\mu} ~~~{\rm at}~~X \to -\infty \; . \label{lim3}
 \end{equation}

One has to understand the limit $|X| \to \infty $ in the following
sense: it describes the system behavior close to the transition
point but outside  the fluctuation region. For the quantum gas
($mU \ll 1$) it means $1 \ll |X| \ll 1/mU $. Of course, one may
easily calculate system properties for any $|X| \gg 1$ using MF
theory presented above, and, take care of the phonon contribution
to the non-quasicondensate and superfluid densities at $ n \sim
T/U $ [contained in the integrals of Eqs.~(\ref{rel7}) and
(\ref{normal})], or, instead of Eq.~(\ref{rel10}), consider a more
accurate expression for the dilute density limit $ n \approx -
(mT/2\pi)   \ln \big[ 1-e^{-\theta m U} \big] $  to include the
Boltzmann gas into the picture. Such obvious generalizations are
not considered in this paper.

\noindent \underline{The vicinity of the Kosterlitz-Thouless point}.
The thermodynamic limit close to the point of the
Berezinskii-Kosterlitz-Thouless transition requires simulations
of exponentially large systems, and thus can hardly be treated
numerically without renormalization group (RG) analysis of finite
size corrections. Fortunately, one can take advantage of the
Kosterlitz-Thouless RG equations that
describe the flow of the superfluid density $n_s(L)$ with
increasing the system size $L$.
In terms of dimensionless function $f_L=(\pi/2mT) \, n_s(L)$
these equations are \cite{KT}
\begin{equation}
{{\rm d} f_L \over {\rm d} \ln L} \, = \, -y^2 f_L^2 \; ,
\label{KT1}
\end{equation}
\begin{equation}
{{\rm d} y \over {\rm d} \ln L} \, = \,  2(1-f_L) \, y \; ,
\label{KT2}
\end{equation}
where $y(L)$ is the vortex-pair fugacity. By excluding
variable $y$ and integrating the remaining RG equation,
one obtains the following  relation
\begin{equation}
F(f_{L_2}, f_{L_1}, \kappa) \, = \, 4 \ln(L_2/L_1) \; ,
\label{RG}
\end{equation}
where $F$ is defined as an integral
\begin{equation}
F(a,b,\kappa )\, = \, \int_a^b {{\rm d} t \over t^2(\ln t -
\kappa) +t} \; , \label{F}
\end{equation}
and $\kappa (X)$ is a size-independent (at $k_cL \gg 1$)
parameter.

By performing large-scale simulations of systems with different
sizes $L_1<L_2<L_3<...$, one may solve Eq.~(\ref{RG}) for parameter
$\kappa (X)$, verify that it is system size independent, and
then determine the thermodynamic value $f(X) = f_{L=\infty}(X)$
from the relation
\begin{equation}
1/f + \ln f = \kappa \; ,
\label{ff}
\end{equation}
that immediately follows from (\ref{RG})-(\ref{F}) at $L \to
\infty$. Since Eq.~(\ref{ff}) has a root only at $\kappa \geq 1$,
we conclude that $\kappa = 1$ corresponds to the critical point.
In contrast to the superfluid density, $\kappa(X)$ has no singularities
at the critical point $X=0$ and may be expanded into Taylor series
\begin{equation}
\kappa (X) \approx 1+\kappa ' X + ... \;.
\label{kappa}
\end{equation}
The solution of Eq.~(\ref{ff})
for small $X$ is then given by formula (\ref{K}).

\section{Numerical procedure}
\label{sec:numerical_procedure}
Although all derivations presented in the previous section were
done for the quantum BG, we expect them to be universal and apply
for any model with effective long wave-length Hamiltonian
(\ref{psi4}) with small $mU$. Classical lattice algorithms are
much more efficient than quantum ones and allow high-accuracy
simulations of very large system sizes. Also, simulations of the
classical lattice model directly test the idea of universality,
since they have to agree with all
Eqs.~(\ref{rel1}),~(\ref{lim2}),~(\ref{lim1}),~(\ref{f_g}),~
(\ref{lim3}).

Our simulations were done for the simple square lattice
Hamiltonian
\begin{equation}
H = \sum_{{\bf k} \in BZ} [E({\bf k}) - \mu ] |\psi_k |^2 +
 {U\over 2 } \sum_i |\psi_i|^4\; ,
\label{H_lat}
\end{equation}
where $\psi_k$ is the Fourier transform of the lattice field
$\psi_i$, and $E({\bf k}) = [ 2-  \cos (k_xa) -  \cos (k_ya)
]/ma^2$ is the tight-binding dispersion law;  momentum ${\bf k}$
being defined within the first Brillouin zone (BZ). We employed
the recently developed Worm algorithm for classical statistical
systems \cite{Worm} which has direct Monte Carlo estimators for
all quantities of interest here and does not suffer from critical
slowing down.

We performed simulations for system sizes $L=64$, $128$, $256$,
$512$ and two values of interaction strength $U=1/4$ and $U=1/16$
to eliminate finite-size and finite-$U$ corrections to the
results. Each quantity for each point in $X$ was calculated with
relative accuracy better then $10^{-3}$ (down to $10^{-4}$ for
smaller system sizes). Our final results for $\theta (X)$, $f(X)$,
and $g(X)$ presented in Figs. \ref{fig1}, \ref{fig2}, and
\ref{fig4} below are thus obtained with accuracy better than
$1~\%$ (the largest error bars of order $ 1~\%$ are in the
vicinity of the critical point where finite-size corrections are
the largest; the errorbars for large $|X|$ shrink down to $0.3
\div 0.1~\%$). Error bars are shown in all plots but typically
they are much smaller than the point size. All the relevant data
are mentioned in Table \ref{tab:table1}. In Figures
\ref{fig1}-\ref{fig4} the MC data are presented by dots and
compared with the asymptotic analytic solutions shown by lines.

\section{Simulation results}
\label{sec:comparison}.
\begin{figure}[tbp]
\includegraphics[width=6.5cm]{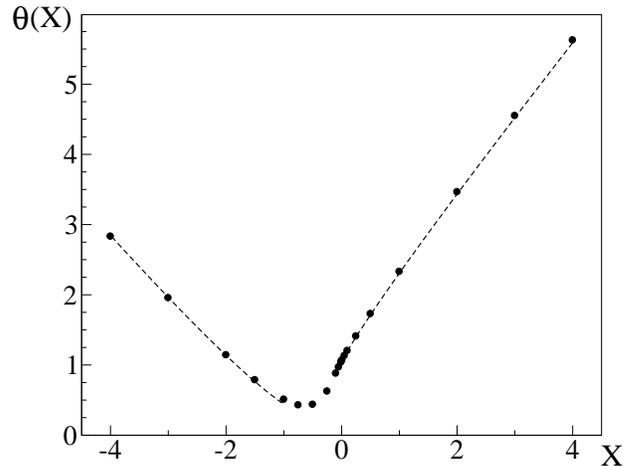}
\vspace*{-2.cm} \caption{The Equation of state $\theta (X)$
compared to its asymptotic large $|X|$ expressions, see
Eqs.~(\ref{lim1}) and (\ref{lim3})
}
\label{fig1}
\end{figure}
\begin{figure}[tbp]
\includegraphics[width=6.5cm]{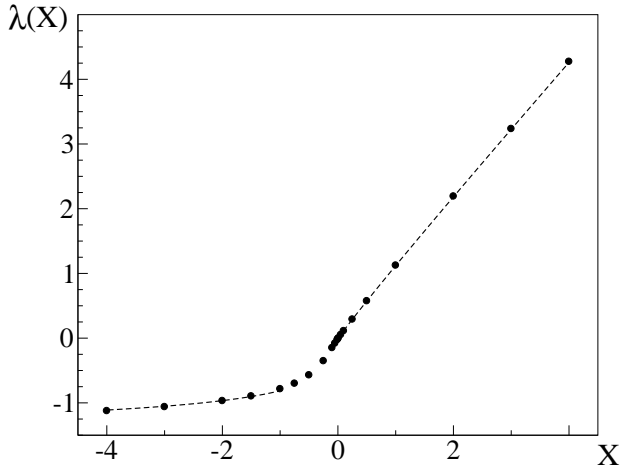}
\vspace*{-2.cm} \caption{The $\lambda (X)$ function compared to
its asymptotic large $|X|$ expressions} \label{fig1a}
\end{figure}

\begin{figure}[tbp]
\includegraphics[width=6.5cm]{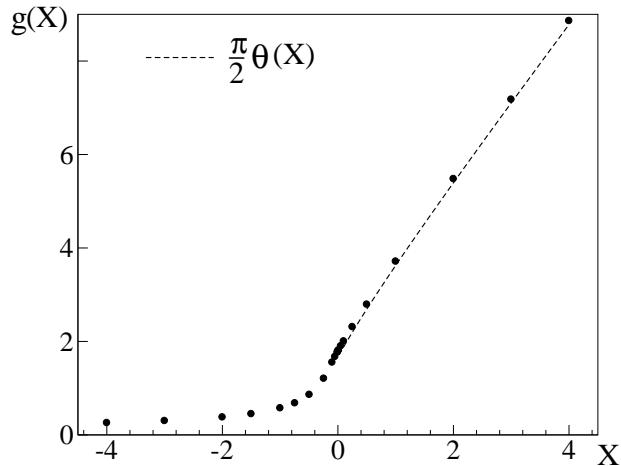}
\vspace*{-2.cm} \caption{The quasicondensate density dependence on
$X$ compared to the asymptotic large $|X|$ behavior predicted by
Eq.~(\ref{lim2})} \label{fig2}

\end{figure}

\begin{figure}[tbp]
\includegraphics[width=6.5cm]{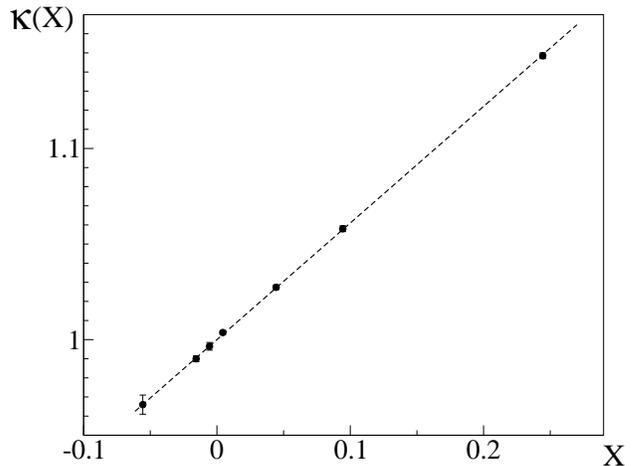}
\vspace*{-2.cm}
\caption{RG parameter $\kappa (X)$ obtained from finite-size scaling
of the data according to Eqs.~(\ref{RG}) and (\ref{F}) and fitted
using linear in $X$ expansion with $d\kappa /dX = 0.61$ }
\label{fig3}
\end{figure}

\begin{figure}[tbp]
\includegraphics[width=6.5cm]{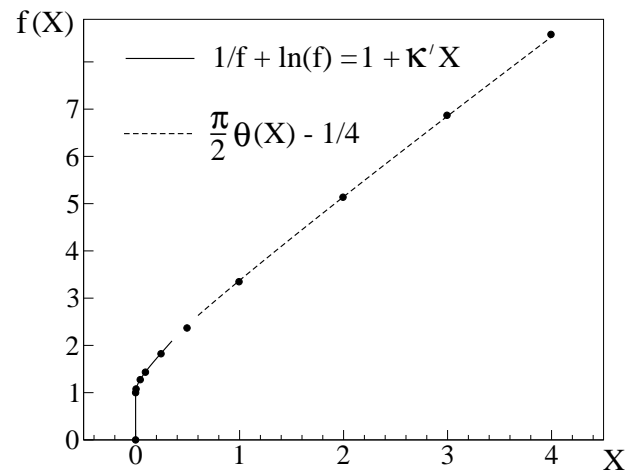}
\vspace*{-2.cm}
\caption{The superfluid density dependence on $X$ compared to the
asymptotic large $|X|$ behavior according to Eq.~(\ref{f_g}),
and the Kosterlitz-Thouless theory for $X \le 0.25 $
according to Eq.~(\ref{ff})}
\label{fig4}
\end{figure}

In Figs.~\ref{fig1},~\ref{fig1a},~\ref{fig2}, and \ref{fig4} we show our final
results for the scaling functions. To compare them with
analytic predictions we first determine the value of the $\theta_0$
parameter from the $\theta(x)$ plot to verify that it agrees
with the result predicted by Eq.~(\ref{rel1}). We find that
\begin{equation}
\theta_0 = 1.068 \pm 0.01 \;,
\label{theta0new}
\end{equation}
thus confirming the universality of parameter $\theta_0$.
Knowing $\xi_\mu$ is all we need to handle the asymptotic behavior at
large $|X|$. The best fits correspond to $\xi_{\mu} = 13.4$,
which, within the error bars, coincides with Eq.~(\ref{xi_mu}).
The agreement between the data and asymptotic laws
Eqs.~(\ref{lim1}) and (\ref{lim3}) in Figs.~\ref{fig1} and
~\ref{fig1a} is remarkable.

The same is also true for the quasicondensate density
(see Fig.~\ref{fig2}). The data perfectly agree with the idea of
quasicondensate which is not entirely obvious at a first glance
for the system without long-range order. We thus confirm that
$n_0$, rigorously defined through the correlation function
Eq.~(\ref{Q}), plays the same role as the genuine condensate density in
the 3D theory.

As explained above, the analysis of the superfluid density data
near the critical point is based on the Kosterlitz-Thouless RG
equations. We found that for $X\in (-0.1,0.25)$ the data scale
according to the RG Eqs.~(\ref{RG}) and (\ref{F}) with negligible
finite-$U$ corrections, and $\kappa (X)$ dependence is well
described by the linear in $X$ expansion Eq.~(\ref{kappa}), see
Fig.~\ref{fig3}. Having determined the first derivative of
$\kappa$ at the critical point as
\begin{equation}
\kappa ' = {d\kappa \over dX} \bigg| _{X=0} = 0.61 \pm 0.01 \; ,
\label{kappa'}
\end{equation}
we proceed with the solution of the thermodynamic value of
$f(X)$ using Eq.~(\ref{ff}). The results are plotted in Fig.~\ref{fig4}
along with the asymptotic large $X$ law
$f(X) \to (\pi /2)\, \theta (X) -1/4$. The two limits match almost
perfectly around $X=0.5$ and describe all the data points with exceptional
accuracy.

\section{Conclusions}
\label{sec:conclusion}.
\begin{figure}[tbp]
\includegraphics[width=6.5cm]{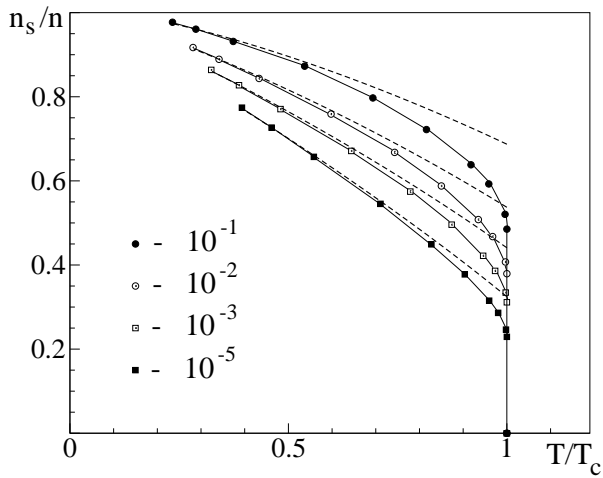}
\vspace*{-2.cm}
\caption{Temperature dependence
of $n_s/n$ for small $mU=10^{-1}$, $10^{-2}$, $10^{-3}$,
$10^{-5}$. Points are connected with lines to guide
the eye.  The dashed lines are the mean-field theory results
with exact relation between $T_c$ and particle density }
\label{fig5}
\end{figure}

\begin{figure}[tbp]
\includegraphics[width=6.5cm]{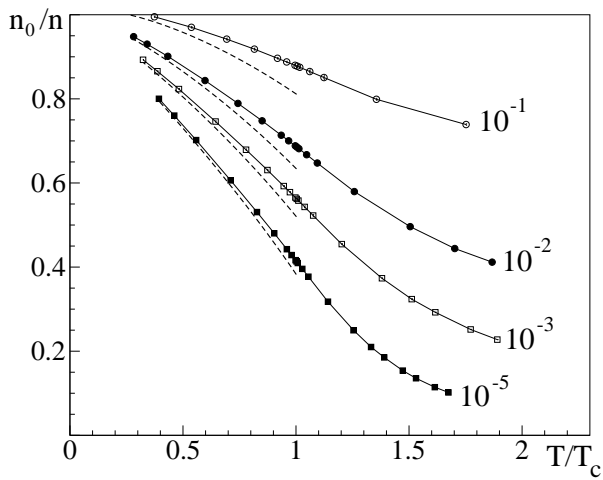}
\vspace*{-2.cm}
\caption{Temperature dependence
of $n_0/n$ for small $mU=10^{-1}$, $10^{-2}$, $10^{-3}$,
$10^{-5}$. Points are connected with lines to guide the eye.
The dashed lines are the mean-field theory results }
\label{fig6}
\end{figure}

\begin{figure}[tbp]
\includegraphics[width=6.5cm]{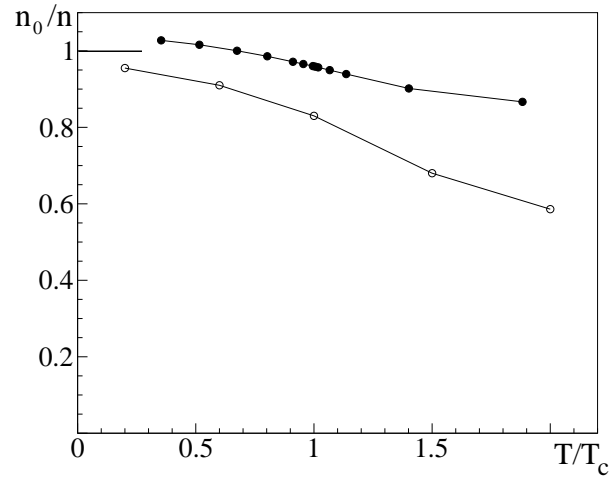}
\vspace*{-2.cm}
\caption{Temperature dependence
of $n_0/n$ for $mU=0.2$ derived from  universal relations
 (filled circles) and simulated for the quantum lattice BG
(open circles). Points are connected
with lines to guide the eye }
\label{fig7}
\end{figure}

The most striking result of this study is that the Equation of
state, $\theta (X)$, and the quasicondensate density, $g(X)$, are
predicted by the MF theory with accuracy of few percent all the
way to the critical point. Even at the critical point the
difference between the data and asymptotic relations is barely
visible. This outcome is quite unexpected because the superfluid
density does show deviations from the MF theory for $X<0.5$.

Hence, from the side of the superfluid phase, the boundary of the
fluctuation region corresponds to $X \sim 0.5$. To express this
estimate in terms of temperature (density), we relate $X$ and $T$
at a constant density ($X$ and $n$ at a constant temperature).
Equation (\ref{dn}) with $n_c$ from Eq.~(\ref{xi_n}) may be
written as
\[
{mT \over n} ={ 2 \pi  \over \ln (  \xi /mU ) + 2\pi \lambda (X) }
\;,
\]
or
\begin{equation}
 {T_c(n) \over T}= {n \over n_c(T)}  = 1+ {2\pi \over \ln ( \xi /mU ) }
 ~ \lambda(X)
\; , \label{T}
\end{equation}
and maps the control parameter $X$ onto $T/T_c(n)$ or $n/n_c(T)$.
This relation is not universal because  $T_c$, $n_c$, and the
r.h.s. of Eq.~(\ref{T}) depend on the ultra-violet cutoff. For
$X=0.5$ and, say, $mU=0.1$ we find that the fluctuation region is
roughly $(T_c-T)/T_c \sim (n-n_c)/n_c \sim 0.3$.

It is instructive to compare results for the superfluid density of
the weakly interacting  BG in the more conventional $n_s/n$ {\it
vs} $T/T_c$ plot at a constant particle density which may be
immediately obtained from the universal relations. The superfluid
density is given by
\begin{equation}
{ n_s \over n } = {2mT \over \pi n}\, f(X) \equiv {4 ( T/T_c )
\over  \ln (  \xi /mU ) } \, f(X) \; , \label{nsT}
\end{equation}
which along with Eq.~(\ref{T}) defines a parametric dependence of
$n_s/n$ on $T/T_c$. One may construct then a modified MF theory
result by substituting $f(X)$ in this relation with $\pi \theta
(X)/2-1/4$, see Eq.~(\ref{f_g}), and $\theta (X)$ from
Eq.~(\ref{lim1}). An analogous expression for $n_0$ is obtained by
replacing $f(X)$ with $g(X)$.

In Fig.~\ref{fig5} we show the comparison between the MF
solution and the data for small effective interaction $mU=10^{-1}, ~
10^{-2}, ~10^{-3}, ~10^{-5}$.
We see that the fluctuation region is still of order $T_c$
even for $mU = 0.01$ --- clearly that small coupling parameter
may not be obtained by going to the dilute limit for the gas
of hard-core particles and thus
any realistic discussion of the experimental data should involve
the proper description of the fluctuation region.
Even for $mU=0.01$ the familiar MF
formula $n_s/n \approx 1-T/T_c$ does not work at all --- the
slope keeps changing with $T$ and the left-most slope does not
point to $n_s/n=1$. However, if one uses an exact relation between
$T_c$ and $n$, then the MF approach described above is capable of
reproducing the data for $X>0.5$.

The other remark concerns the asymmetry of the fluctuation region
at $T<T_c$ and $T>T_c$.  The minimum in the Fig.~\ref{fig1} plot
for $\theta(X)$ and the agreement with the MF laws already suggest
that the fluctuation region is much broader on the normal side.
When MF results for $n_s (T)$ in Fig. \ref{fig5} are extrapolated
to higher temperatures, the intersection with the temperature axis
is still at $ \approx 1.5~T_c$  for $mU$ as small as  $\sim
10^{-5}$! We draw the same conclusion from the $n_0/n$ plot shown
in Fig.~\ref{fig6}: The decay of the quasicondensate density $n_0
\equiv \sqrt{Q}$, which is a measure of local non-Gaussian
correlations, is quite extended into the normal state.

Our universal relations are obtained in the limit $mU \ll 1$. From
a practical point of view it is important to estimate a typical
value of $mU$ at which  higher-order corrections to our results
become unimportant. To this end we note that if we plot the
quasicondensate density as a function of $T/T_c$  for various
values of $mU$, see Fig.~\ref{fig6} we will find that for $mU=0.1$
the ratio $n_0/n$ exceeds unity already at $T \approx 0.4~T_c$.
This unphysical result tells us that non-universal corrections for
the quantum BG are {\it not} negligible even for $mU \sim 0.1$ (at
least for the quasicondensate density; for numerical reasons they
might be smaller for other quantities). We thus expect that
universal expressions established in this study are likely to work
without limitations only for $mU$ significantly smaller than
$0.1$. To understand the situation with the quantum corrections
quantitatively, we compare in Fig. \ref{fig7} our results for
$n_0/n$ at $mU=0.2$ with the previously reported \cite{Kagan}
results for the quantum lattice model \cite{rem2}. The comparison
suggests that at $T \sim T_c$ the quantum correction to the
quasicondensate density is $\sim mU$ (in relative units). The sign
of the correction is negative, that is we are dealing with a
quantum depletion of the quasicondensate.

For helium films on various substrates \cite{Reppy,Hallock}, and
spin-polarized atomic hydrogen on helium film \cite{Safonov}, the
value of $mU$ is most probably of order unity \cite{KSS}.

In the recently created quasi-2D gas of sodium atoms
\cite{Ketterle}, $mV_0$ is of order $10^{-2}$, and this system is
supposed to be described by our results very precisely. In
experiments with {\it trapped} gases the quantity directly
relevant to the experimental setup is $\lambda(X)$ since it
describes, according to Eq.~(\ref{dn}), the density profile in the
trapping potential which is smooth enough to guarantee the
hydrostatic regime. In this regime the density variation over the
mode-coupling radius $r_c \sim 1/k_c$ [the data of
Ref.~\onlinecite{PRS} suggest $r_c \approx 2/m(UT)^{1/2}$ ] is
small, and the coordinate dependence of density reduces to $n
\equiv n(T, \mu({\bf r}))$, where $n(T,\mu)$ is the homogeneous
equation of state, $\mu({\bf r}) = V_{\rm ext} ({\bf r}) + {\rm
const}$, and $V_{\rm ext}$ is the trapping potential. It follows
from Fig.~\ref{fig6} that for $mU \sim 10^{-2}$ the size of the
fluctuation region on the normal side is of order unity. Thus when
the density at the trap center is tuned to the critical point,
practically the whole density profile finds itself in the
fluctuation region where MF equations do not work.

\section{Acknowledgments}

This work was supported by the National Science Foundation under
Grant DMR-0071767 and NASA. BVS acknowledges a support from Russian
Foundation for Basic Research under Grant 01-02-16508, from the
Netherlands Organization for Scientific Research (NWO), and from
the European Community under Grant INTAS-2001-2344.

\begin{table}
\caption{\label{tab:table1} Final results (after taking care of
finite-size and finite-$U$ corrections) for the scaling functions
$\theta (X)$, $g(X)$, and $f(X)$. }
\begin{ruledtabular}
\begin{tabular}{llll}
 ~~~~$X  $ ~~~~& ~~~~$\theta (X)$~~~~ & ~~~~$g(X)$~~~~ & ~~~~$f(X)$~~~~  \\ \hline
 -4.0056 &      2.8363(3)  &  0.2657(3)  &  ~    \\
 -3.0056 &      1.9603(5)  &  0.3094(4)  &  ~   \\
 -2.0056 &      1.1472(6)  &  0.3866(6)  &  ~   \\
 -1.5056 &      0.791(1)   &  0.4561(8)  &  ~    \\
 -1.0056 &      0.514(1)   &  0.581(1)   &  ~     \\
 -0.7556  &      0.434(3)   &  0.688(3)   &  ~      \\
 -0.5056  &      0.442(3)   &  0.869(2)   &  ~     \\
 -0.2556  &      0.630(4)   &  1.214(4)   &  ~      \\
 -0.1056  &      0.885(9)   &  1.560(7)   &  ~     \\
 -0.0556  &      0.973(9)   &  1.680(7)   &  ~    \\
 -0.0156  &      1.041(9)   &  1.774(8)   &   ~  \\
 -0.0056  &      1.061(4)   &  1.800(3)   &   ~   \\
~0.0044  &      1.075(5)   &  1.821(4)   &  1.077(5)    \\
~0.0444&      1.137(5)   &  1.908(4)   &  1.274(4)   \\
~0.0944
&1.208(5)   &  2.013(4)   &  1.433(5)   \\
~0.2444& 1.415(6) &
2.319(6)   &  1.823(6)   \\ ~0.4944& 1.734(6)   & 2.800(7) &
2.37(1)     \\ ~0.9944& 2.334(9) &3.721(8)   &  3.348(6)     \\
   ~1.9944 &      3.469(9)   &  5.47(1)    &  5.135(10)    \\
  ~2.9944 &      4.554(9)   &  7.19(1)    &  6.87(1)    \\
  ~3.9944 &      5.631(8)   &  8.870(6)   &  8.58(1)    \\
\end{tabular}
\end{ruledtabular}
\end{table}


\begin{thebibliography}{99}
\bibitem{rem} Only leading order terms in $mU$ are universal
for any weakly interacting $|\psi|^4$ model. Higher-order
corrections are, generally speaking, model-dependent.
\bibitem{rem2} The correlator $K_2$ of Ref.~\onlinecite{Kagan}
relates to our correlator $Q$ by the formula
$Q=n^2(2-\tilde{K}_2)$, where $\tilde{K}_2(n)=K_2(n)/K_2(n \to
0)$.
\end{thebibliography}
\end{document}